\documentclass[reprint,superscriptaddress,groupedaddress,
amsmath,amssymb,aps,pra,longbibliography,floatfix,]{revtex4-1}

\usepackage{times}
\usepackage[colorlinks=true,linkcolor=blue,urlcolor=blue,citecolor=blue]{hyperref}
\usepackage{graphicx}
\usepackage{bm}
\usepackage{bbold}
\usepackage{dsfont}
\usepackage{verbatim}
\usepackage{xcolor}
\usepackage{mathtools}

\usepackage{scalerel}
\usepackage{tikz}
\usetikzlibrary{svg.path}
\definecolor{orcidlogocol}{HTML}{A6CE39}
\tikzset{
	orcidlogo/.pic={
		\fill[orcidlogocol] svg{M256,128c0,70.7-57.3,128-128,128C57.3,256,0,198.7,0,128C0,57.3,57.3,0,128,0C198.7,0,256,57.3,256,128z};
		\fill[white] svg{M86.3,186.2H70.9V79.1h15.4v48.4V186.2z}
		svg{M108.9,79.1h41.6c39.6,0,57,28.3,57,53.6c0,27.5-21.5,53.6-56.8,53.6h-41.8V79.1z M124.3,172.4h24.5c34.9,0,42.9-26.5,42.9-39.7c0-21.5-13.7-39.7-43.7-39.7h-23.7V172.4z}
		svg{M88.7,56.8c0,5.5-4.5,10.1-10.1,10.1c-5.6,0-10.1-4.6-10.1-10.1c0-5.6,4.5-10.1,10.1-10.1C84.2,46.7,88.7,51.3,88.7,56.8z};
	}
}
\newcommand\orcid[1]{\href{https://orcid.org/#1}{\mbox{\scalerel*{
				\begin{tikzpicture}[yscale=-1,transform shape]
				\pic{orcidlogo};
				\end{tikzpicture}
			}{R}}}}

\renewcommand{\Re}{{\rm Re}}
\renewcommand{\Im}{{\rm Im}}
\newcommand{\Tr}{{\rm Tr}}
\newcommand{\cA}{{\mathcal A}}
\newcommand{\cS}{{\mathcal S}}
\newcommand{\cC}{{\mathcal C}}
\newcommand{\cD}{{\mathcal D}}
\newcommand{\cL}{{\mathcal L}}

\newcommand{\cU}{{\mathcal U}}
\newcommand{\cV}{{\mathcal V}}
\newcommand{\nbar}{\bar{n}}
\newcommand{\bt}{\boldsymbol{\tau}}
\newcommand{\bx}{\boldsymbol{\xi}}
\newcommand{\be}{\boldsymbol{\epsilon}}
\newcommand{\bp}{\boldsymbol{\psi}}

\begin{document}
	\title{Pulsed characteristic-function measurement of a thermalizing harmonic oscillator}
	
	\author{Ralf Betzholz\,\orcid{0000-0003-2570-7267}} 
	
	\author{Yu Liu\,\orcid{0000-0001-7251-2310}}
	\email{yuliu\_phys@hust.edu.cn}
	
	\author{Jianming Cai}
	\affiliation{School of Physics, International Joint Laboratory on Quantum Sensing and Quantum Metrology, Institute for Quantum Science and Engineering, Huazhong University of Science and Technology, Wuhan 430074, China}
	
	\date{\today}

	\begin{abstract}
		We present a method for the direct measurement of the Wigner characteristic function of a thermalizing harmonic oscillator that is completely inaccessible for control or measurement. The strategy employs a recently proposed probe-measurement-based scheme [\href{https://journals.aps.org/prl/abstract/10.1103/PhysRevLett.122.110406}{Phys. Rev. Lett. \textbf{122}, 110406 (2019)}] which relies on the pulsed control of a two-level probe. We generalize this scheme to the case of a nonunitary time evolution of the target harmonic oscillator, describing its thermalization through contact to a finite-temperature environment, given in the form of a Lindblad master equation. This generalization is achieved using a superoperator formalism and yields analytical expressions for the direct measurement of the characteristic function, accounting for the decoherence during the measurement process. 
	\end{abstract}
	
	
	\maketitle

	\section{Introduction}
	\label{sec:intro}
	The exact knowledge of the quantum state of a system is indispensable for most modern
	quantum experiments. This includes not only fundamental tests of quantum mechanics but also the rapidly evolving field of quantum-information processing where also continuous-variable quantum systems are of great interest~\cite{Braunstein2005}. For continuous-variable states, the quantum state can not only be represented by a density operator but also by an equivalent description in terms of phase-space distributions or characteristic functions~\cite{Cahill1969a,Cahill1969b}. In order to obtain the knowledge of these phase-space distributions, a standard tool is continuous-variable quantum-state tomography~\cite{Lvovsky2009}, for example, using homodyne detection~\cite{Vogel1989,Leonhardt1997} or photon-number parity measurements~\cite{Banaszek1996,Nehra2020}.
	
	However, most tomography methods require measurement access to the continuous-variable degree of freedom. The situation changes appreciably in systems where such a direct access is not attainable. In this kind of scenario, an established method is to employ a readily accessible measurement probe~\cite{Lutterbach1997,Tufarelli2011,Tufarelli2012,Yazdanpanah2017,Paz-Silva2019,Liu2019,Yang2019}, often given by two-level quantum systems. For example, such a procedure has been employed in a well-known probe-based direct measurement of the Wigner function~\cite{Lutterbach1997}.
	
	In this paper, we propose a strategy for a direct probe-based measurement of the Wigner characteristic function~\cite{Scully1997,Carmichael2002,Glauber2007} of an inaccessible harmonic oscillator where the probe is formed by a two-level quantum system that exerts a state-dependent force on the harmonic oscillator. The measurement is direct in the sense that the values of the distribution can be directly connected to  probe-measurement outcomes, without the necessity of subsequent numerical processing, such as the inverse Radon transform~\cite{Radon1917,Vogel1989,Schleich2001}. Nevertheless, in order to obtain the Wigner function itself, or other phase-space distributions, one would still have to perform the complex Fourier transform of the measured characteristic function.  

	Contrary to previous probe-based schemes for the direct measurement of phase-space distributions, such as the Wigner function, the method for the measurement of the characteristic function presented here does not require the control of the harmonic oscillator, such as a displacement operation prior to the measurement procedure~\cite{Leibfried1996,Kim1998,Bertet2002}, the controllability of the coupling to the probe~\cite{Casanova2012,Taketani2014,Fluhmann2019}, or an additional third degree of freedom~\cite{Singh2010}. In principle, displacement operations by coherent fields do not necessarily pose an obstacle. Our aim here is to present an alternative, for example, for situations when they are not applicable. Where in our alternative the characteristic function is measured, which, being the Fourier transform of the Wigner function, contains the same information.
	
	Explicitly, we use the interference generated by a pulse sequence applied to the probe in order to engineer the time evolution of the elements of the density operator that are off-diagonal in the probe degree of freedom in order. These off-diagonal elements can then be directly connected to values of the Wigner characteristic function at desired points in reciprocal phase space. We generalize a recently proposed pulsed reconstruction~\cite{Liu2019} to a sequence of temporally nonequidistant pulses applied to the probe, resulting in the possibility for a better reconstruction of the quantum state of the target harmonic oscillator. We additionally account for decoherence of the harmonic oscillator, described using a Lindblad master equation for a Markovian thermalization by a finite-temperature environment~\cite{Scully1997,Carmichael2002}. This makes it possible to compensate for such a thermalization that occurs during the measurement process~\cite{Moya-Cessa1999a,Moya-Cessa1999b,Juarez-Amaro2003,Deglise2008} and, nevertheless, obtain the exact values of the characteristic function of the initial state. In order to do so, we employ a superoperator formalism to derive the analytical expressions for the time evolution under the applied pulse sequence and show how the probe-measurement outcomes are related to the values of the characteristic function. Furthermore, the pulse sequence applied to the probe is of such a form that it makes the quantum-state reconstruction robust against noise acting on the probe two-level system during the measurement process. This is achieved by incorporating pulsed dynamical decoupling~\cite{Viola1999,Lange2010,Naydenov2011} in the form of a spin-echo method~\cite{Levitt2008} where the probe states are flipped at appropriate times and the action of noise with a long correlation time can be reverted.
	
	The paper is structured as follows. In Sec.~\ref{sec:sequence} we introduce the thermalization dynamics of the harmonic oscillator and the probe under the pulse sequence that is applied to the probe. The resulting effective propagators of the initial density operator of the harmonic oscillator are derived in Sec.~\ref{sec:propagators}. These propagators are then used in Sec.~\ref{sec:characteristic} to derive which values of the characteristic function can be obtained, how they are related to the probe-sequence parameters, and how they are connected to expectation values of probe measurements. Section~\ref{sec:examples} shows some explicit examples, before we summarize our results and draw conclusions in Sec.~\ref{sec:conclusion}. Moreover, for self-consistency, useful identities and technical details (e.g., various superoperator and phase-space-distribution identities) supporting the calculations in the main text are presented and derived in the Appendices~\ref{app:displacements}--\ref{app:cummuting}.

	\section{System and pulse sequence}
	\label{sec:sequence}
	In this section, we introduce the general form of the dissipative dynamics and the pulse sequence that we employ. The composite system we investigate consists of the thermalizing harmonic oscillator, whose quantum state we want to reconstruct, and a two-level probe that is accessible for a control through pulses as well as measurements.
	
	\subsection{Thermalizing harmonic oscillator}
	\label{sec:sequence.1}
	The thermalizing harmonic oscillator we consider has the frequency $\nu$, whereas its annihilation and creation operators are denoted by $a$ and $a^\dagger$, respectively. Its unitary evolution is thereby determined by the Hamiltonian $H=\hbar\nu (a^\dagger a+1/2)$. Besides the unitary dynamics, generated by this Hamiltonian, the description of the free time evolution of its density operator $\rho$ also includes a nonunitary contribution originating in the interaction of the harmonic oscillator with its thermal environment at the reservoir temperature $\Theta$. This interaction can be characterized by the damping rate $\gamma$ and the mean thermal occupation $\bar{n}=[\exp(\hbar\nu/k_B\Theta)-1]^{-1}$. 
	In a Born-Markov approximation, this leads to the Lindblad master equation $\partial\rho/\partial t=\cL\rho$ in which the action of the Liouvillian superoperator $\cL$, that generates the dynamics of $\rho$, is given by~\cite{Scully1997,Carmichael2002}
	\begin{gather}
	\cL\rho=-i\nu [a^\dagger a,\rho]+\frac{(\nbar+1)\gamma}{2}\big(2a\rho a^\dagger-\{a^\dagger a,\rho\}\big)\nonumber\\
	\label{eq:L}
	+\frac{\nbar\gamma}{2}\big(2a^\dagger\rho a-\{aa^\dagger,\rho\}\big),
	\end{gather}
	with the commutator $[\,\cdot\,,\,\cdot\,]$ and the anticommutator $\{\,\cdot\,,\,\cdot\,\}$. Here, and throughout the subsequent text, we represent superoperators acting on the space of harmonic-oscillator operators by calligraphic letters and, in case they have a dependence on parameters, we write their arguments in square brackets.
	
	\subsection{Two-level probe and state-dependent interaction}
	\label{sec:sequence.2}
	As a measurement probe, we consider a generic two-level system described by the Hamiltonian $H_\text{probe}=(\hbar\omega/2)\sigma_z$, where $\sigma_\kappa$ for $\kappa=x,y,z$ denote the usual Pauli operators. We represent the eigenstates of $\sigma_z$ to the eigenvalues $\pm1$ by $|\pm\rangle$. 
	The $\pm$ notation, that is often used for the $\sigma_x$ eigenstates, is chosen over common notations for the $\sigma_z$ eigenstates, such as $\uparrow\downarrow$, for later notational convenience. 
	
	We assume the interaction between the probe and the harmonic oscillator to be of the form
	\begin{equation}
	\label{eq:H_int}
	H_{\rm int}=\frac{\hbar g}{2}\sigma_z(a+a^\dagger),
	\end{equation}
	i.e., a probe-state-dependent constant force acting on the harmonic oscillator~\cite{Agarwal2012,Liu2019} with the same strength but opposite signs depending on the probe state. 
	
	Such a coupling can be realized in a variety of electromechanical systems, for example, when an electronic spin is coupled to the motional degree of freedom of a mechanical element emanating a magnetic field or is embedded in a mechanical oscillator subject to a spatially inhomogeneous magnetic field~\cite{Rabl2009,Peng2016,Cao2017}. Also, such an interaction can be realized in harmonically trapped particles with an internal electronic degree of freedom, such as ions in a linear Paul trap. In this scenario, the above interaction can be realized with a magnetic field gradient along the trap axis~\cite{Johanning2009,Khromova2012,Sriarunothai2017}, i.e., a magnetic field of the form $B(z)=B_0+B_1z$, where the $z$ direction is the trap axis. With the position operator $z\sim a+a^\dagger$ of the ion, this leads to a coupling to $\sigma_z$ of the form~\eqref{eq:H_int}. Alternatively, it could be implemented by appropriately engineered laser configurations, which then additionally allows to conveniently turn the interaction on and off on demand. More details on the feasibility of our method in such experimental realizations have been discussed in Ref.~\cite{Liu2019} for the case of the harmonic oscillator with thermalization. Cooling and heating of trapped ions can be well described by a master equation of the form~\eqref{eq:L}~\cite{Lindberg1984,Turchette2000a,Turchette2000b}.
	
	\subsection{Pulse sequence}
	\label{sec:sequence.3}
	In order to apply an indirect control on the otherwise inaccessible harmonic oscillator, the probe two-level system is initially prepared in a known pure state, given by the state vector $|\bp\rangle=\psi_+|+\rangle+\psi_-|-\rangle$, whose probability amplitudes $\bp=(\psi_+,\psi_-)$ fulfill $|\psi_+|^2+|\psi_-|^2=1$. The initially separable density operator of the combined system, before the pulse sequence we introduce below is applied, therefore, has the form
	\begin{equation}
	\label{eq:initial_state}
	\varrho(0)=|\bp\rangle\langle\bp|\,\rho_0.
	\end{equation}
	Here, and in the following, we denote by $\rho_0$ the unknown density operator of the harmonic oscillator at the initial time $t=0$, namely, the state we want to reconstruct. After the initialization, we apply a sequence of $2N$ pulses on the probe that each exactly flip its states according to $|\pm\rangle\longleftrightarrow|\mp\rangle$. 
	
	These so-called $\pi$ pulses are assumed to be instantaneous, and we consider a sequence of free-evolution times between the pulses where always pairs of two adjacent free-evolution times are of the same duration $\tau_n$, with $n=1,\dots,N$. This means one can represent the entire sequence as
	\begin{equation*}
	[\pi-\tau_N-\pi-\tau_N]-\dots-[\pi-\tau_2-\pi-\tau_2]-[\pi-\tau_1-\pi-\tau_1],
	\end{equation*}
	where the time line goes from right to left, such that it corresponds to the order of the individual time-evolution operators that are applied. Here, every $\pi$ stands for an instantaneous $\pi$ pulse and the square brackets indicate the $N$ individual pulse-sequence segments of length $2\tau_n$, for $n=1,\dots,N$. The total length of the sequence is thereby given by $T=2\sum_{n=1}^N\tau_n$. The $N$ values of the free evolution times are the quantities we will control and vary in order to measure the quantum state of the harmonic oscillator and we summarize them in the vector $\boldsymbol{\tau}=(\tau_1,\dots,\tau_N)$. 
	
	Since the interaction commutes with the probe Hamiltonian, in the following, we will always work in the interaction picture with respect to $H_\text{probe}$. The dynamics of the full system's density operator $\varrho$ then contains a coherent contribution from the interaction, that is, a commutator with $H_{\rm int}$ from Eq.~\eqref{eq:H_int} and the thermalization of the harmonic oscillator described by the superoperator $\cL$ from Eq.~\eqref{eq:L}. Explicitly, during the free evolution it is governed by the master equation
	\begin{equation}
	\label{eq:master}
	\frac{\partial}{\partial t}\varrho=\cL\varrho-\frac{ig}{2}[\sigma_z(a+a^\dagger),\varrho].
	\end{equation}
	Here, we mention that in such a description we have made the assumption that the thermalization of the harmonic oscillator is not affected by the interaction to the probe. This assumption is widely used in quantum-optical calculations, where damping terms of the Lindblad form are phenomenologically added despite the presence of interaction terms, which is a good description if one is not deep in the strong-coupling regime. With the initial density operator $\varrho(0)$ from Eq.~\eqref{eq:initial_state} this linear differential equation can be integrated until time $t$, yielding
	\begin{align}
	\varrho(t)&=\Big(|\psi_+|^2|+\rangle\langle +|\,e^{\cC_+t}+|\psi_-|^2|-\rangle\langle -|\,e^{\cC_-t}\nonumber\\
	&+\psi_+\psi_-^\ast|+\rangle\langle -|\,e^{\cA_+t}+\psi_-\psi_+^\ast |-\rangle\langle +|\,e^{\cA_-t}\Big)\rho_0.
	\end{align}
	In this expression, the reduced propagators that act on the initial harmonic-oscillator density operator $\rho_0$ are given in terms of the superoperators $\cC_{\pm}$ and $\cA_{\pm}$, whose action is 
	\begin{gather}
	\label{eq:commutator}
	\cC_{\pm}\rho=\cL\rho\mp \frac{ig}{2}[a+a^\dagger,\rho],\\
	\label{eq:anticommutator}
	\cA_{\pm}\rho=\cL\rho\mp \frac{ig}{2}\{a+a^\dagger,\rho\}.
	\end{gather}
	They, respectively, include a commutator and an anticommutator with the dimensionless position operator $a+a^\dagger$, that originate from the interaction~\eqref{eq:H_int} with the probe. 
	
	After every free-evolution time, the probe states are flipped through the $\pi$ pulses and the dynamics of the density operator can be integrated over the subsequent free-evolution time, during which the signs in the superoperators $\cC_{\pm}$ and $\cA_{\pm}$ are changed. Iterating this procedure for the full pulse sequence then results in the final state
	\begin{align}
	\label{eq:full_state}
	\varrho(\bt)&=\Big(|\psi_+|^2|+\rangle\langle +|\,\cU_+[\bt]+|\psi_-|^2|-\rangle\langle -|\,\cU_-[\bt]\nonumber\\
	+&\psi_+\psi_-^\ast|+\rangle\langle -|\,\cV_+[\bt]+\psi_-\psi_+^\ast|-\rangle\langle +|\,\cV_-[\bt]\Big)\rho_0.
	\end{align}
	This is the density operator after the total length $T$ of the entire pulse sequence, however, since it naturally depends on the choice of the free-evolution times $\bt$, in the remainder of the text, we explicitly specify this dependence in the notation $\varrho(\bt)$. In the above expression for the state of the total system, we have introduced the two pairs of effective harmonic-oscillator propagators
	\begin{gather}
	\label{eq:propagator_diagonal}
	\cU_\pm[\bt]=\prod_{n=1}^Ne^{\cC_\mp\tau_n}e^{\cC_\pm\tau_n},\\
	\label{eq:propagator_offdiagonal}
	\cV_\pm[\bt]=\prod_{n=1}^Ne^{\cA_\mp\tau_n}e^{\cA_\pm\tau_n},
	\end{gather}
	which also depend parametrically on $\bt$. Here, and in the following, we always use the superoperator product convention $\prod_{n=1}^N\mathcal{X}_n=\mathcal{X}_N\dotsb\mathcal{X}_2\mathcal{X}_1$, corresponding to the sequence in which the propagators for the different building blocks of the pulse sequence are applied to the initial density operator. These superoperators describe the effective time evolution of the harmonic oscillator and the influence, including the interference described by $\cV_{\pm}[\bt]$, of the applied pulse sequence.

	\section{Dynamics under the pulse sequence}
	\label{sec:propagators}
	In this section, we will derive the explicit form of the four effective reduced propagators $\cU_\pm[\bt]$ and $\cV_\pm[\bt]$ and their action on $\rho_0$. The first kind of propagators $\cU_\pm[\bt]$ are, in fact, not strictly relevant for the later quantum-state reconstruction scheme. However, we, nevertheless, show the derivation of their explicit form briefly, for the sake of completeness and because this derivation is instructive for the subsequent treatment of $\cV_{\pm}[\bt]$. The second kind of propagators $\cV_\pm[\bt]$ do not constitute trace-preserving maps since they merely describe the interference generated by the pulse sequence.
	
	Many of the following derivations greatly rely on some of the well-known properties of the displacement operators $D(\alpha)=\exp(\alpha a^\dagger-\alpha^\ast a)$, such as their quasiclosure under multiplication. Here, we introduce two kinds of displacement superoperators, which we will refer to as \textit{superdisplacements}. Writing the dynamics in term of these superdisplacements makes it possible to conveniently trace back the actions of $\cC_{\pm}$ and $\cA_{\pm}$ to the one of the Liouvillian superoperator $\cL$ introduced in Eq.~\eqref{eq:L}. This is a great advantage since the properties of $\cL$, such as its diagonalization~\cite{Briegel1993,Englert2002} or the corresponding Green's function in phase space~\cite{Carmichael2002}, have been well studied. In this sense, we will simply find transformations that transform these superoperators into one with well-known properties. When dealing with operators, one usually finds unitary transformations that transform some operator into another one with a known eigenbasis. Here, our treatment of superoperators is analogous, we merely take the procedure to the level of superoperators.

	\subsection{Diagonal part: $\cU_\pm$}
	\label{sec:propagators.1}
	For the treatment of the superoperators on the diagonal in the probe-state basis it is convenient to introduce the symmetric superdisplacement $\cD[\epsilon]$, whose action is given by
	\begin{equation}
	\cD[\epsilon]\rho=D(\epsilon)\rho D^\dagger(\epsilon).
	\end{equation}
	From this definition it can be seen that the multiplication property of the displacement operators directly translates to the superoperator form $\cD[\epsilon]\cD[\varsigma]=\cD[\epsilon+\varsigma]$, i.e., the superdisplacements are closed under multiplication, giving them a genuine group character, contrary to the displacement operators themselves where an additional phase factor arises during the multiplication. Analogous to the displacement operators they fulfill $\cD[\epsilon]\cD[-\epsilon]=\mathcal{I}$ with the identity superoperator $\mathcal{I}$. 
	
	We now employ this supertransformation with an appropriate displacement parameter $\varepsilon$ so that we can bring the action of $\cC_{\pm}$ back to the one of $\cL$, viz., 
	\begin{equation}
	\label{eq:DLD}
	\cC_\pm =\cD[\mp\varepsilon]\cL\cD[\pm\varepsilon].
	\end{equation}
	Here, the appropriate choice for the superdisplacement parameter $\varepsilon$, such that this condition is fulfilled is given by the relative coupling strength
	\begin{equation}
	\varepsilon=\frac{g}{2\tilde{\nu}},
	\end{equation}
	which we defined, for convenience and brevity, in terms of the complex frequency
	\begin{equation}
	\tilde{\nu}=\nu-\frac{i\gamma}{2}.
	\end{equation}
	A detailed derivation of this choice of $\varepsilon$ is shown in Appendix~\ref{app:displacements.1}. Considering the superoperator defined in Eq.~\eqref{eq:commutator}, one sees that this transform essentially transforms a thermalizing harmonic oscillator under a constant force into a simple thermalizing harmonic oscillator. The transform~\eqref{eq:DLD} can be written in an exponential form according to
	\begin{equation}
	\label{eq:symmetric_displacement}
	e^{\cC_\pm t}=\cD[\mp\varepsilon]e^{\cL t}\cD[\pm\varepsilon],
	\end{equation}
	which is convenient for the derivation of the explicit form of the propagators $\cU_\pm[\bt]$ defined in Eq.~\eqref{eq:propagator_diagonal}. 
	
	At this point we want to move the free-thermalization propagator $\exp(\cL t)$ to the very right and, therefore, make use of the identity
	\begin{equation}
	\label{eq:switch_relation_1}
	e^{\cL t}\cD[\varepsilon]=\cD\left[\varepsilon e^{-i\tilde{\nu} t}\right]e^{\cL t}.
	\end{equation}
	This relation has already been used in other scenarios for the direct measurement of phase-space distribution under the influence of dissipation~\cite{Deglise2008}, and a detailed derivation of it is presented in Appendix~\ref{app:cummuting}. Using the multiplication property of the superdisplacements, Eq.~\eqref{eq:symmetric_displacement} can be directly rearranged according to
	\begin{equation}
	\label{eq:expC}
	e^{\cC_\pm t}=\cD\left[\mp\varepsilon\left(1-e^{-i\tilde{\nu}t}\right)\right]e^{\cL t},
	\end{equation}
	where we find that $\exp(\cC_{\pm} t)$ may be written as a free thermalization followed by a time-dependent superdisplacement. 
	
	We can now successively apply this relation to bring the propagator $\cU_\pm[\bt]$ for the full pulse sequence in a similar form. Equation~\eqref{eq:propagator_diagonal} for the full propagation of length $T$ thereby becomes
	\begin{equation}
	\cU_\pm[\bt]=\prod_{n=1}^N\cD\left[\pm\varepsilon\left(1-e^{-i\tilde{\nu}\tau_n}\right)^2\right]e^{2\cL\tau_n}.
	\end{equation}
	Repeatedly applying identity~\eqref{eq:switch_relation_1} in order to bring all free-thermalization propagators to the very right and contracting all super\-displacements into a single one finally yields
	\begin{equation}
	\cU_\pm[\bt]=\cD[\pm\upsilon(\bt)]e^{\cL T},
	\end{equation}
	with the function 
	\begin{equation}
	\label{eq:ut}
	\upsilon(\bt)= \varepsilon e^{-i\tilde{\nu}T}\sum_{n=1}^N\left(1-e^{-i\tilde{\nu}\tau_n}\right)^2
	\prod_{k=1}^n e^{2i\tilde{\nu}\tau_k},
	\end{equation}
	which can be readily shown by induction. This results shows that the propagation under the full pulse sequence can be written as a free thermalization followed by a single superdisplacement, whose displacement arguments $\pm\upsilon(\bt)$ depend on the vector $\bt$ of free-evolution times.
	
	From the formal expression~\eqref{eq:full_state} for the time-evolved state $\varrho(\bt)$ of the full system, one can perform the partial trace over the probe degree of freedom to obtain the reduced state of the harmonic oscillator, denoted by $\rho(\bt)\equiv\Tr_\text{probe}\{\varrho(\bt)\}$. Employing the above expressions, we can write
	\begin{equation}
	\rho(\bt)=\mathcal{P}[\bp,\bt]e^{\cL T}\rho_0,
	\end{equation}
	where the effective action of the pulse sequence, which is applied to the probe, on the harmonic oscillator is given by the superoperator
	\begin{equation}
	\mathcal{P}[\bp,\bt]=|\psi_+|^2\cD[\upsilon(\bt)]+|\psi_-|^2\cD[-\upsilon(\bt)].
	\end{equation}
	This superoperator naturally depends parametrically on the initial probe-state probability amplitudes $\bp$ as well as the free-evolution times $\bt$, and it is applied to the free thermalization of the initial state, given by $\exp(\cL T)\rho_0$. Not surprisingly, the action is that of effective probe-state dependent superdisplacements with opposite signs of the displacement parameter, whose weight is given by the population probabilities of the initial probe state.
	
	\subsection{Off-diagonal part: $\cV_\pm$}
	\label{sec:propagators.2}
	It is actually the off-diagonal propagators $\cV_{\pm}[\bt]$, containing the anticommutators $\cA_{\pm}$ in which we are mainly interested since they describe the interference generated by the pulse sequence. Although the procedure in this case is slightly more intricate, it follows along similar lines as the preceding one for the diagonal-term propagators $\cU_\pm[\bt]$ involving $\cC_{\pm}$. The difference is that here we have to employ an asymmetric or as we will call it \textit{skew} superdisplacement. A transform of this kind has already been employed in the diagonalization of superoperators containing an anticommutator with the dimensionless position operator~\cite{Betzholz2014,Torres2019,Betzholz2020}. Using the abbreviation $\bx=(\xi_1,\xi_2,\xi_3)$ for three displacement parameters we have to employ in this case, we define the skew superdisplacement as
	\begin{equation}
	\cS[\bx]\rho=e^{\xi_1 a}D(\xi_2)\rho D^\dagger(\xi_3)e^{-\xi_1 a},
	\end{equation}
	instead of the previously introduced symmetric superdisplacements $\cD[\varepsilon]$. This makes the following calculation somewhat more tedious, however, still relatively straightforward, keeping in mind the above derivation for the symmetric case. At this point we mention that one could define an even more general supertransform in terms of four displacement parameters. However, since the above definition, with merely three parameters, is sufficient for what follows we abstain from doing so. The product property for these skew superdisplacements reads
	\begin{equation}
	\label{eq:SpSm}
	\cS[\boldsymbol{\epsilon}]\cS[\boldsymbol{\varsigma}]=e^{(\epsilon_3-\epsilon_2)\varsigma_1}e^{\frac{1}{2}(\epsilon_2\varsigma_2^\ast-\epsilon_2^\ast\varsigma_2-\epsilon_3\varsigma_3^\ast+\epsilon_3^\ast\varsigma_3)}\cS[\boldsymbol{\epsilon}+\boldsymbol{\varsigma}],
	\end{equation}
	which is straightforward to derive from the multiplication property of the displacement operators. We briefly note two more of their properties, namely, that they are related to the identity superoperator by
	$\cS[\bx]\cS[-\bx]=\exp(\xi_1(\xi_2-\xi_3))\mathcal{I}$ 
	and that the symmetric superdisplacements are the special case $\cS[\boldsymbol{\xi}]=\cD[\varepsilon]$ for $\boldsymbol{\xi}=(0,\varepsilon,\varepsilon)$.
	
	The action of $\cA_{\pm}$ can also be reduced to the one of $\cL$ according to
	\begin{equation}
	\label{eq:SLS}
	\cA_{\pm}=e^{\xi_1(\xi_3-\xi_2)}\cS[\mp\bx]\cL\cS[\pm\bx]-\Gamma\mathcal{I}.
	\end{equation}
	The appropriate displacement parameters one has to choose are derived in Appendix~\ref{app:displacements.2} and read
	\begin{equation}
	\label{eq:xi}
	\bx=\frac{g}{4|\tilde{\nu}|^2}
	\begin{pmatrix}
	-4i(2\nbar+1)\gamma\\
	2\nu+i(4\bar{n}+1)\gamma\\
	-2\nu+i(4\bar{n}+3)\gamma
	\end{pmatrix},
	\end{equation}
	whereas the constant multiplying the identity superoperator is given by
	\begin{equation}
	\label{eq:decay_constant}
	\Gamma=\frac{g^2(2\nbar+1)\gamma}{2|\tilde{\nu}|^2}.
	\end{equation}
	This also can be written in its exponential form
	\begin{equation}
	\label{eq:asymmetric_displacement}
	e^{\cA_\pm t}=e^{-\Gamma t}e^{\xi_1(\xi_3-\xi_2)}\cS[\mp\bx]e^{\cL t}\cS[\pm\bx],
	\end{equation}
	where, contrary to the symmetric superdisplacement transform from Eq.~\eqref{eq:symmetric_displacement}, there appears an exponential decay in Eq.~\eqref{eq:asymmetric_displacement} with the decay rate $\Gamma$, which is identical for both signs in $\cA_{\pm}$ and is likewise derived in Appendix~\ref{app:displacements.2}. This constant $\Gamma$ is a positive real number, leading to a strict decay of the interference described by $\cV_{\pm}[\bt]$. Furthermore, it goes to zero if either $\gamma$ or $g$ vanish, whereas at zero temperature it is still finite. 
	
	The time evolution of the terms that are off-diagonal in the probe-state basis are governed by the superoperators $\cV_\pm[\bt]$ introduced in Eq.~\eqref{eq:propagator_offdiagonal}. In that form, it is already decomposed into a product accounting for the evolution in the separate pulse-sequence blocks labeled by $n=1,\dots,N$. Using the reduction of $\exp(\cA_{\pm} t)$ to the action of skew superdisplacements $\cS[\pm\bx]$ and the free thermalization $\exp(\cL t)$, given by Eq.~\eqref{eq:asymmetric_displacement}, as well as the product property~\eqref{eq:SpSm} of these skew superdisplacements, we can rewrite the generator of the evolution of the off-diagonal parts for a single pulse-sequence block as
	\begin{align}
	\label{eq:V}
	e^{\cA_\mp t}e^{\cA_\pm t}=&e^{-2\Gamma t}e^{3\xi_1(\xi_3-\xi_2)}\nonumber\\
	&\times\cS[\pm\bx]e^{\cL t}\cS[\mp 2\bx]e^{\cL t}\cS[\pm\bx].
	\end{align}
	This is the expression that we will employ in the following. It will turn out that this form is convenient for the direct measurement of the characteristic function we introduce below, which relies on the expansion of the initial density operator in the basis formed by the displacement operators. It is convenient since the action of both the skew superdisplacements and the free thermalization maps displacement operators onto displacement operators.

	\section{Characteristic-function measurement}
	\label{sec:characteristic}
	Having the explicit expression for the time evolution of the initial state at hand, which was derived in the previous section, we now come to the core of this paper, namely, how measurements of the probe can be connected to values of the Wigner characteristic function of the harmonic oscillator. We mention again, that contrary to many tomographic methods, the measurement of the characteristic function is direct in the sense that the outcomes of the probe measurements can be directly converted into values of the characteristic function~\cite{Tufarelli2011} without further numerical processing.
	
	\subsection{Wigner characteristic function}
	\label{sec:characteristic.1}
	In the following, we will find that it is advantageous to write the dynamics of the initial state $\rho_0$ in terms of its Wigner characteristic function~\cite{Scully1997,Carmichael2002,Glauber2007}. Before we do so, we first give a brief reminder on the definition of this function and some of its properties. 
	
	The completeness of the displacement operators~\footnote{
		The displacement operators $D(\beta)$ form a complete set of operators, such that every bounded operator $X$ has the representation $X=\int d^2\beta\,\Tr\{D(\beta)X\}D^\dagger(\beta)/\pi$ (q.v. Ref.~\cite{Cahill1969a} for a proof).
	} 
	makes it possible to write the harmonic-oscillator density operator as
	\begin{equation}
	\label{eq:displacement_expansion}
	\rho_0=\frac{1}{\pi}\int d^2\beta\,\chi(\beta)D^\dagger(\beta),
	\end{equation}
	with the Wigner characteristic function $\chi(\beta)$, defined as the expectation value of a displacement operator $D(\beta)$, viz.,
	\begin{equation}
	\label{eq:characteristic_function}
	\chi(\beta)=\Tr\{D(\beta)\rho_0\}.
	\end{equation}
	This function on reciprocal phase space is, in fact, nothing but the complex Fourier transform of the celebrated Wigner function~\cite{Wigner1932} and thereby contains the same information~\footnote{
		The well-known Wigner function $W(\alpha)$ is the complex Fourier transform of the characteristic function $\chi(\beta)$, given by $W(\alpha)=\int d^2\beta\,\chi(\beta)\exp(\alpha\beta^\ast-\alpha^\ast\beta)/\pi^2$ (q.v. Refs.~\cite{Cahill1969a,Cahill1969b})
	}.
	The unit-trace condition, which translates to a unit square integrability of the Wigner distribution, is reflected in the fact that $\chi(0)=1$. Furthermore, the Hermiticity of the density operator and the property $D(-\beta)=D^\dagger(\beta)$ directly yield $\chi(-\beta)=\chi(\beta)^\ast$. This implies that the knowledge of the characteristic function on a half-plane of the complex numbers representing the reciprocal phase space is sufficient to fully reconstruct $\rho_0$. 
	
	\subsection{Dynamics in terms of displacement operators}
	The linearity of the superoperators used in the above formulation of the dynamics implies that, in a characteristic-function description, one needs to obtain their action on the displacement operator $D^\dagger(\beta)$ appearing in the expansion~\eqref{eq:displacement_expansion}. Especially, we will see that it is the terms in Eq.~\eqref{eq:full_state} which are off-diagonal in the probe-state basis carrying the information on the harmonic-oscillator density operator that can be extracted through probe measurements. Keeping in mind the expansion~\eqref{eq:displacement_expansion}, this means we need to derive the expressions $\cV_\pm[\bt]D^\dagger(\beta)$, which will be performed in the remainder of this section. The harmonic-oscillator state $\rho_0$ evolved under the action of these superoperators is then given by the integral over those expressions weighted with the initial characteristic function $\chi(\beta)$. 
	
	Examining Eq.~\eqref{eq:V}, one finds that there are two main ingredients necessary for the derivation of $\cV_\pm[\bt]D^\dagger(\beta)$, namely, the action of: (i) the skew superdisplacement $\cS[\be]$ and (ii) the free-thermalization propagator $\exp(\cL t)$ on a displacement operator $D(\varsigma)$. Let us start with (i), which can easily be derived by applying the multiplication property of displacement operators twice and, furthermore, using $\exp(\epsilon a)a^\dagger\exp(-\epsilon a)=a^\dagger+\epsilon$, resulting in
	\begin{gather}
	\cS[\be]D(\varsigma)=e^{\epsilon_1(\epsilon_2-\epsilon_3)}e^{\frac{1}{2}(\epsilon_2^\ast\epsilon_3-\epsilon_2\epsilon_3^\ast)}e^{\frac{1}{2}[(\epsilon_2+\epsilon_3)\varsigma^\ast-(\epsilon_2^\ast+\epsilon_3^\ast)\varsigma]}\nonumber\\
	\times e^{\epsilon_1\varsigma}D(\varsigma+\epsilon_2-\epsilon_3).\label{eq:cSD}
	\end{gather}
	A detailed derivation of (ii) can be found in Appendix~\ref{app:damped_displacements}, yielding the action of the free-thermalization propagator $\exp(\cL t)$ on a displacement operator, given by
	\begin{equation}
	\label{eq:LD}
	e^{\cL t}D(\varsigma)=e^{\gamma t}e^{-\nbar(t)|\varsigma(t)|^2}D(\varsigma(t)),
	\end{equation}
	where we introduced the abbreviation $\varsigma(t)=\varsigma\exp(-i\tilde{\nu}^\ast t)$ and the time-dependent thermal occupation
	\begin{equation}
	\nbar(t)=\left(\nbar+\frac{1}{2}\right)\left(1-e^{-\gamma t}\right).
	\end{equation}
	Here, we find the important property that we have mentioned before, namely, that both of these superoperators map displacement operators onto displacement operators, apart from some prefactors. It is exactly this fact that makes our method for the direct measurement of the characteristic function possible.
	
	With these two properties at hand, one can successively apply Eqs.~\eqref{eq:cSD} and~\eqref{eq:LD} in order to derive the action of a single sequence-segment superoperator defined in Eq.~\eqref{eq:V} on an arbitrary displacement operator $D(\varsigma)$. This is a straightforward, yet somewhat cumbersome, calculation that yields the identity
	\begin{align}
	e^{\cA_\mp t}e^{\cA_\pm t}D(\varsigma)&=e^{2(\gamma-\Gamma)t+\frac{\Gamma\gamma}{|\tilde{\nu}|^2}+f(t,\pm\varsigma)}\nonumber\\
	\times &D\left(\varsigma e^{-2i\tilde{\nu}^\ast t}\pm 2\varepsilon^\ast \left(1-e^{-i\tilde{\nu}^\ast t}\right)^2\right),
	\end{align}
	with the function
	\begin{gather}
	f(t,\varsigma)=2\Gamma\,\Im\left\{\frac{1}{\tilde{\nu}^\ast}\left(2-e^{-i\tilde{\nu}^\ast t}\right)^2
	+\frac{\varsigma}{g}\left(1-e^{-i\tilde{\nu}^\ast t}\right)^2\right\}\nonumber\\
	-\bar{n}(t)e^{\gamma t}\Bigg[\left|2\varepsilon^\ast+\varsigma\right|^2
	+\left|\left(2\varepsilon^\ast+\varsigma\right)e^{-i\tilde{\nu}^\ast t}-4\varepsilon^\ast\right|^2\Bigg].
	\label{eq:phi}
	\end{gather}
	This shows that during every pulse-sequence segment a displacement operator acquires an exponential multiplier and its argument is rotated in addition to a shift proportional to the relative coupling strength, namely, $2\varepsilon^\ast=g/\tilde{\nu}^\ast$. One also finds that the only temperature dependence arises in the exponent through the constant $\Gamma$ and the function $f$.
	
	Let us now come back to the expansion of the initial density operator $\rho_0$ in terms of its characteristic function, i.e., Eq.~\eqref{eq:displacement_expansion}, and extend the technique introduced in this section to the entire pulse sequence. Iterating this procedure for the $N$ pulse-sequence segments yields an expression for $\cV_\pm[\bt]D^\dagger(\beta)$ given by
	\begin{align}
	\label{eq:VD}
	\cV_\pm[\bt]D^\dagger(\beta)=&e^{(\gamma-\Gamma)T+\frac{N\Gamma\gamma}{|\tilde{\nu}|^{2}}+\sum_{n=1}^Nf\left(\tau_n,\beta_\pm^{(n)}\right)}\nonumber\\
	&\times D^{\dagger}\left(e^{-i\tilde{\nu}^\ast T}\left[\beta \mp \zeta(\bt) \right]\right),
	\end{align}
	with the function
	\begin{equation}
	\label{eq:zeta}
	\zeta(\bt)=2\varepsilon^\ast\sum_{n=1}^N\left(1-e^{-i\tilde{\nu}^\ast\tau_n}\right)^2\prod_{k=1}^n e^{2i\tilde{\nu}^\ast\tau_k},
	\end{equation}
	that bears a great resemblance with the function $\upsilon(\bt)$ that occurred in the diagonal case, with a substitution of $\tilde{\nu}$ by its complex conjugate. The exponent in the exponential factor multiplying the resulting displacement operator, in turn, is defined in terms of the displacement parameters $\beta_{\pm}^{(n)}$ given by
	\begin{gather}
	\beta_{\pm}^{(n)}= 2\varepsilon^\ast\sum_{j=1}^{n-1}\left(1-e^{-i\tilde{\nu}^\ast\tau_j}\right)^2\prod_{k=j+1}^{n-1} e^{-2i\tilde{\nu}^\ast\tau_k}\nonumber\\
	\mp\beta\prod_{j=1}^{n-1} e^{-2i\tilde{\nu}^\ast \tau_j}.
	\label{eq:betaN}
	\end{gather}
	
	All the preceding steps have enabled us to finally write out the full expression for the off-diagonal part of the time evolution, which is given by the integral
	\begin{gather}
	\cV_\pm[\bt]\rho_0=\frac{e^{(\gamma-\Gamma)T+\frac{N\Gamma\gamma}{|\tilde{\nu}|^2}}}{\pi}\int d^2\beta\,e^{\sum_{n=1}^Nf\left(\tau_n, \beta_\pm^{(n)}\right)}\chi(\beta)\nonumber\\
	\times D^\dagger\left(e^{-i\tilde{\nu}^\ast T}\left[\beta \mp \zeta(\bt) \right]\right).
	\label{eq:V_charac}
	\end{gather}
	Let us have a closer look at this result. It constitutes the action of the off-diagonal propagator $\cV_\pm[\bt]$ on the initial state in its displacement-operator expansion. We find that the initial characteristic function $\chi(\beta)$, multiplied by an additional exponential, is now the weight function of an expansion in displacement operators shifted by $\mp \zeta(\bt)$ as well as rotated and stretched by the factor $\exp(-i\tilde{\nu}^\ast T)$.

	\subsection{Pauli measurements on the probe}
	\label{sec.4.2}
	Remembering the expression~\eqref{eq:full_state} for the full state $\varrho(\bt)$ of the combined probe and harmonic-oscillator system after the applied pulse sequence, one can now perform measurements on the probe, such as a measurement of the Pauli vector $\boldsymbol{\sigma}=(\sigma_x,\sigma_y,\sigma_z)$ with the outcome $\langle\boldsymbol{\sigma}\rangle_{\bt}=\Tr\{\boldsymbol{\sigma}\varrho(\bt)\}$. It is easily seen that the outcome,
	\begin{equation}
	\langle\boldsymbol{\sigma}\rangle_{\bt}=
	\begin{pmatrix}
	\Tr\{(\psi_+\psi_-^\ast \cV_+[\bt]+\psi_+^\ast \psi_-\cV_-[\bt])\rho_0\}\\
	i\,\Tr\{(\psi_+\psi_-^\ast\cV_+[\bt]-\psi_+^\ast \psi_-\cV_-[\bt])\rho_0\}\\
	|\psi_+|^2-|\psi_-|^2
	\end{pmatrix},
	\end{equation}
	of such a measurement is directly connected with the harmonic-oscillator initial state $\rho_0$. Here, the independence of the $z$ component of the expectation value on the harmonic-oscillator state is due to the trace-preserving nature of the superoperators $\cU_\pm[\bt]$, viz., $\Tr\{\cU_\pm[\bt]\rho_0\}=1$. The $x$ and $y$ components, on the other hand, yield the relation
	\begin{equation}
	\langle\sigma_x\mp i\sigma_y\rangle_{\bt}=2\psi_\pm\psi_\mp^\ast \Tr\{\cV_\pm[\bt]\rho_0\}.
	\end{equation}
	
	The traces can then be evaluated employing the expansion~\eqref{eq:V_charac} of $\cV_\pm[\bt]\rho_0$, which we derived exactly for this purpose. Now, the final step is to employ the property $\Tr\{D(\beta)\}=\pi\delta^2(\beta)$~\cite{Cahill1969a}, with the complex Dirac-$\delta$ distribution $\delta^2(\beta)=\delta(\Re\{\beta\})\delta(\Im\{\beta\})$. The integration over the complex plane can thereby be executed trivially, which finally yields
	\begin{equation}
	\label{eq:main}
	\chi(\pm \zeta(\bt))=C_\pm(\bp,\bt)\langle\sigma_x\mp i\sigma_y\rangle_{\bt},
	\end{equation}
	with the accessible reciprocal-phase-space points $\pm \zeta(\bt)$. The scaling factor is given by
	\begin{equation}
	C_\pm(\bp,\bt)=\frac{e^{\Gamma T-\frac{N\Gamma\gamma}{|\tilde{\nu}|^2}-\sum_{n=1}^Nf\left(\tau_n,\phi_n\right)} }{2\psi_\pm\psi_\mp^\ast},
	\end{equation}
	with
	\begin{equation}
	\label{eq:phin}
	\phi_n=-2\varepsilon^\ast\sum_{j=n}^{N}\left(1-e^{-i\tilde{\nu}^\ast\tau_j}\right)^2\prod_{k=j+1}^{n-1} e^{-2i\tilde{\nu}^\ast\tau_k},
	\end{equation}
	which comes from replacing $\beta$ in Eq.~\eqref{eq:betaN} by $\pm \zeta(\bt)$.
	
	Equation~\eqref{eq:main} constitutes the main result of this paper, because it directly connects expectation values of probe measurements with values of the Wigner characteristic function of the harmonic oscillator undergoing thermalization. By engineering $\bt$, such as to sweep over one-half of the complex plane, in principle, the full characteristic function can be measured directly. The scaling factor $C_\pm(\bp,\bt)$ can be regarded as a weight factor of the characteristic-function measurement. It is only in this factor that $\nbar$ and thereby the temperature appears.

	\section{Examples}
	\label{sec:examples}
	In order to demonstrate how the choice of the pulse sequence influences the points in reciprocal phase space on which the characteristic function can be sampled and how it affects a quantum-state reconstruction, we will show three different examples for the choice of the free-evolution times $\bt$:
	
	(i) A sequence of $N$ equally long free-evolution times~\cite{Liu2019}, according to $\bt=(\tau_0,\dots,\tau_0)$, leading to a total pulse-sequence length $T=2N\tau_0$.
	
	(ii) A sequence with free-evolution times chosen uniformly random with $\bt\in[0,2\pi/\nu]^N$, resulting in an average total evolution time $T=2N\pi/\nu$.
	
	(iii) A sequence of first linearly increasing and then linearly decreasing free-evolution times in the form of  the vector $\bt=(\tau_0,2\tau_0, \dots,(N-1)\tau_0, N\tau_0,(N-1)\tau_0, \dots,2\tau_0,\tau_0)$, yielding a total length $T=2N^2\tau_0$. Since in this case the pulse sequence consists of $2N-1$ segments, one has to replace $N$ by $2N-1$ in the expressions~\eqref{eq:VD}--\eqref{eq:phin}.
	
	For these three pulse sequences, we will first present the points reachable in reciprocal phase space and their dependence on the damping rate $\gamma$. We will furthermore show examples how these points can be employed to measure the characteristic function of some initial quantum states of the harmonic oscillator, which can then be used to infer the density matrix. As a last step, we will analyze the influence of the thermalization on the reconstruction fidelity. 
	
	\subsection{Reachable points in reciprocal phase space}
\begin{figure}[tb]
		\includegraphics[width=\linewidth]{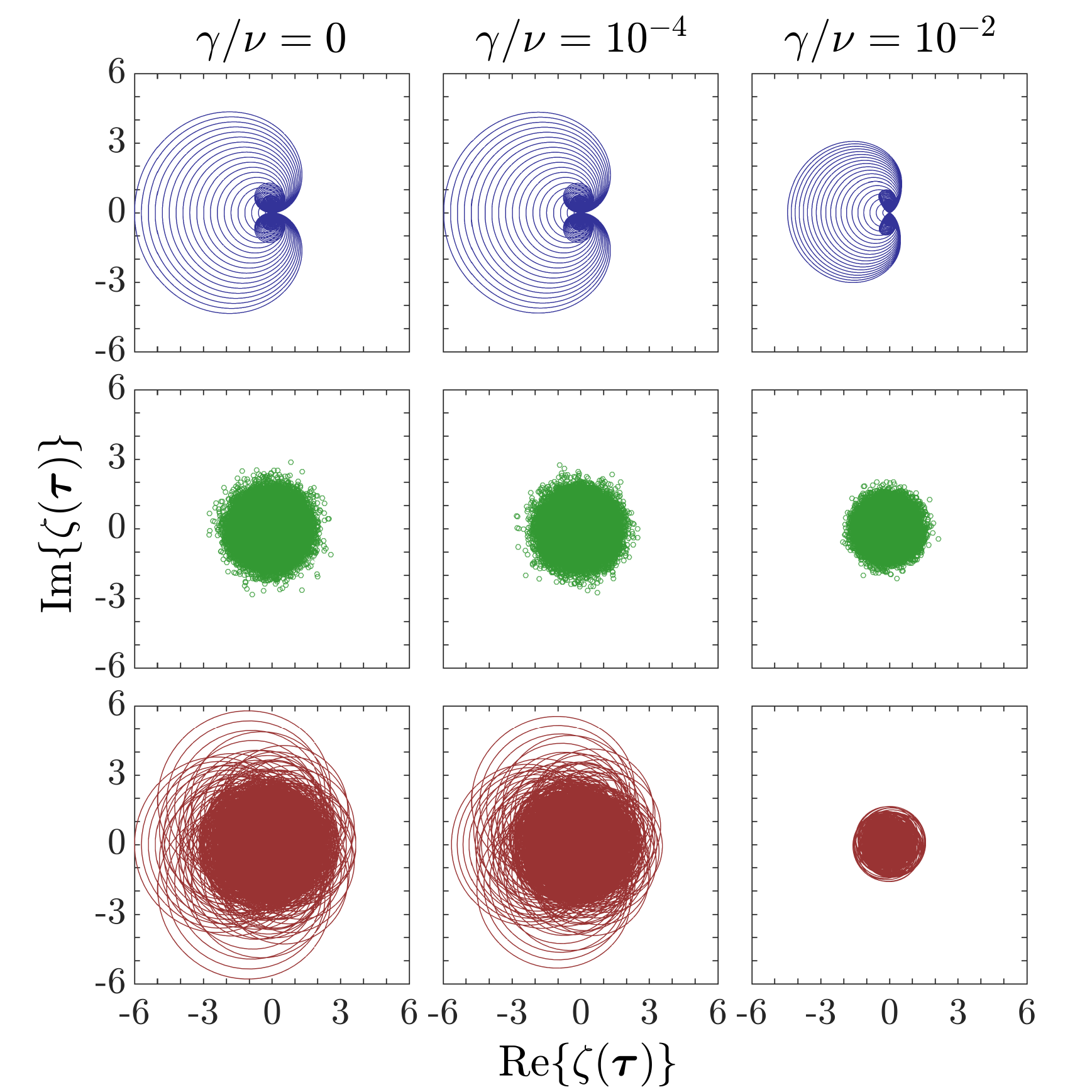}
		\caption{\label{fig:1} Points in reciprocal phase space of the thermalizing harmonic oscillator accessible through probe measurements. The three rows (from top to bottom) correspond to three different choices of the free-evolution times $\bt$: (i) Equidistant pulses (blue), (ii) randomly chosen (green), and (iii) linearly increasing and then decreasing (red) free-evolution times. The three columns correspond to different values of the damping rate: $\gamma/\nu=0,10^{-4},10^{-2}$. The accessible points for all three pulse sequences are plotted for $g/\nu=0.075$ and $N=1,2,\dots,20$. In (i) and (iii)  the curves are for $\tau_0\in[0,2\pi/\nu]$, whereas for (ii) 10\,000 random configurations are shown.} 
	\end{figure}

	Figure~\ref{fig:1} depicts the reachable points $\zeta(\bt)$ in reciprocal phase space for the three pulse sequences (i)--(iii) introduced above and the three damping rates $\gamma/\nu=0,10^{-4},10^{-2}$. The three columns correspond to the three different damping rates, whereas the three rows correspond to the three different pulse sequences. For the coupling strength we have chosen $g/\nu=0.075$ in all cases.
	
	(i) In the first row of Fig.~\ref{fig:1}, which corresponds to the equidistant pulses, every blue curve shows the reachable points $\zeta(\bt)$ for a certain number $N$ of pulses when the free-evolution time $\tau_0$ is swept over the interval $[0,2\pi/\nu]$. The curves from the inside to the outside correspond to the values $N=1,2,\dots,20$. In this case, the expression for the reachable points $\zeta(\bt)$, defined in Eq.~\eqref{eq:zeta}, can be evaluated to
	\begin{equation}
	\zeta(\bt)=4\varepsilon^*\sin(N\nu^*\tau_0)\tan\left(\frac{\nu^*\tau_0}{2}\right)e^{iN\nu^*\tau_0},
	\end{equation} 
	yielding the same result previously obtained for a harmonic oscillator without damping~\cite{Liu2019} if $\nu$ is replaced by the complex frequency $\tilde{\nu}^\ast$. The maximum of the distance to the origin is achieved roughly at $\tau_0=\pi/\nu$. At this free-evolution time, the approximate distance is given by $2g[\exp(-N\pi\gamma/\nu)-2]/\pi\gamma$, which reaches $-4g/\pi\gamma$ when $N$ tends to infinity. For the damping rate $\gamma/\nu=10^{-4}$, no apparent difference to the case without thermalization is visible. For the damping rate $\gamma/\nu=10^{-2}$, however, the curves have shrunk quite substantially, decreasing the area in which the values of the characteristic function can be measured for the same number $N$ of pulses.
	
	(ii) In the second row of Fig.~\ref{fig:1}, the green points represent reciprocal-phase-space points generated using 10\,000 random pulse sequences for $N=1,2,\dots,20$. In this case, the points are centered densely around the origin, and the impact of the increasing damping rate is less pronounced.
	
	(iii) The last row in Fig.~\ref{fig:1} depicts the linear pulse sequence. The different red curves correspond to $N=1,2,\dots,20$, when $\tau_0\in[0,2\pi/\nu]$. Similar to the equidistant pulse sequence, the maximum distance to the origin is approximately reached for $\tau_0=\pi/\nu$. However, compared to (ii) more points are found in the area around the origin. It is also visible that in this case the thermalization of the harmonic oscillator has a strong impact on the accessible points. Even for $\gamma/\nu=10^{-4}$ a small contraction can be seen, whereas for $\gamma/\nu=10^{-2}$ the area is immensely reduced. This is due to the fact, that the linear pulse sequence has an appreciably longer total length scaling quadratically with $N$ over which the thermalization occurs.

	\subsection{Characteristic-function measurement}
\begin{figure}[tb]
		\includegraphics[width=\linewidth]{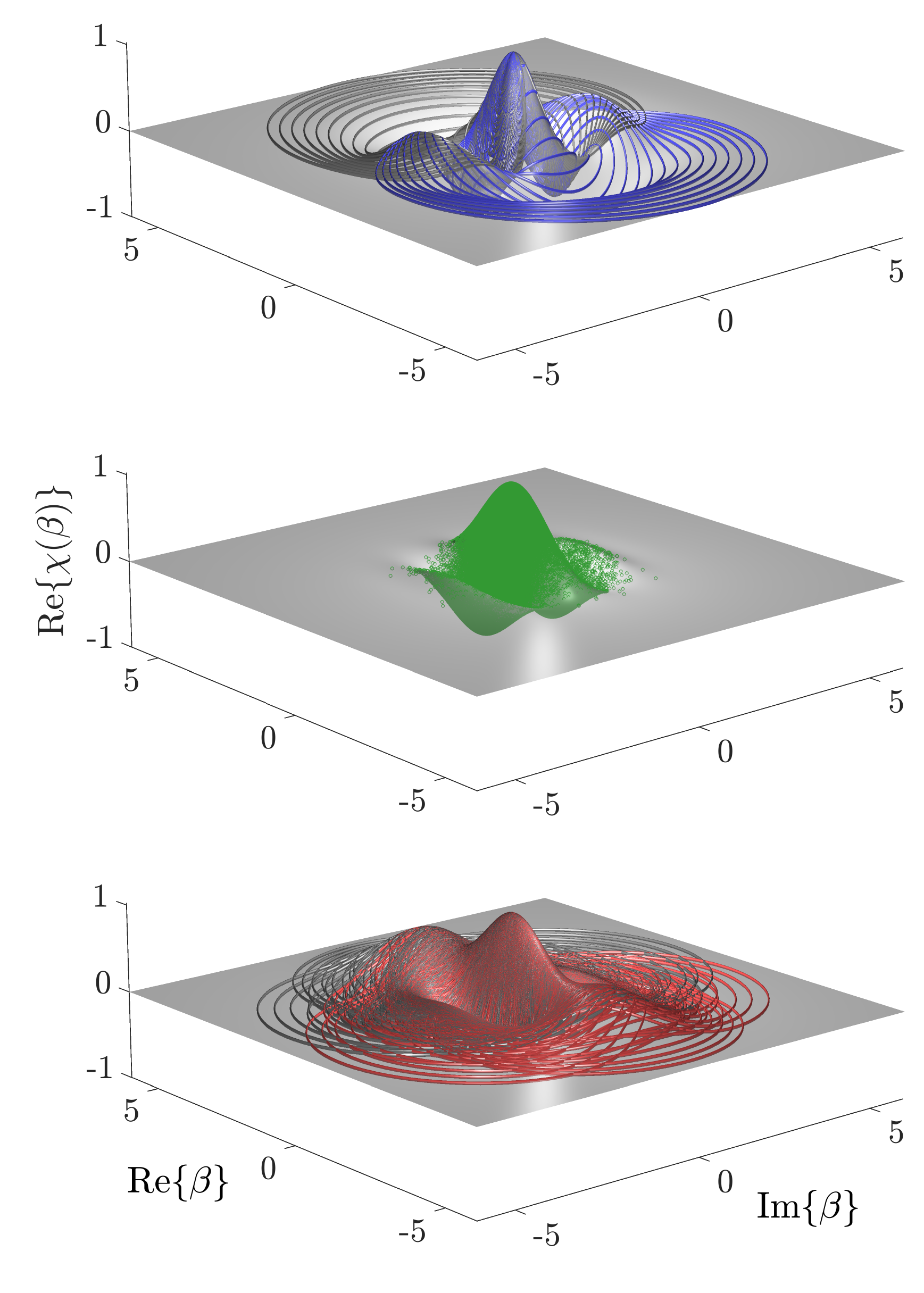}
		\caption{\label{fig:2} Illustration of the values of the Wigner characteristic function of a thermalizing harmonic oscillator that can be measured using the pulse sequences (i)--(iii) for different initial states $\rho_0$: An equal superposition of the Fock sates $|1\rangle$ and $|3\rangle$ for the equidistant pulse sequence (i) [upper panel], the coherent state $D(3/2)|0\rangle$ for the random pulse sequence (ii) [central panel], and a cat state $\propto [D(3/2)+D(-3/2)]|0\rangle$ for the linear pulse sequence (iii) [bottom panel]. The gray surfaces show the full characteristic function of the initial states, whereas the curves or points show the values that are accessible through probe measurements with the parameters $g/\nu=0.075$,  $\gamma/\nu=10^{-4}$, and $N=1,2,\dots,20$.} 
	\end{figure}

	In order to demonstrate how the actual measurement of the Wigner characteristic function works, we use the three pulse sequences introduced above and apply them to three initial states of the harmonic oscillator. We choose pure states $\rho_0=\vert\varphi\rangle\langle\varphi\vert$, with a different state vector $\vert\varphi\rangle$ for every pulse sequence. The results are shown in Fig.~\ref{fig:2}, where we used $g/\nu=0.075$, $\gamma/\nu=10^{-4}$, and $N=1,2,\dots,20$ for all three cases, corresponding to the accessible points depicted in the central column of Fig.~\ref{fig:1}.
	
	(i) For the equidistant pulse sequence we choose the initial state $\vert\varphi\rangle = (\vert 1\rangle +\vert 3\rangle)/\sqrt{2}$, i.e., an equal superposition of two Fock states of the harmonic oscillator. In this case, the characteristic function reads~\cite{Cahill1969b}
	\begin{align}
	\chi(\beta)=\frac{e^{-\frac{|\beta|^2}{2}}}{2}\bigg[&L_1^{(0)}\left(|\beta|^2\right)+L_3^{(0)}\left(|\beta|^2\right)\nonumber\\
	&+\frac{1}{\sqrt{6}}\left(\beta^2+\beta^{\ast 2}\right)L_1^{(2)}\left(|\beta|^2\right)\bigg],
	\label{eq:characteristic_Fock}
	\end{align}
	with the Laguerre polynomials $L_n^{(k)}(x)$. In Fig.~\ref{fig:2}, this characteristic function, which is real in this specific case, is shown at the top as a gray surface, and the blue curves show the accessible points $\zeta(\bt)$ where the values can be directly measured using the equidistant pulse sequence. The darker curves show the values that can be obtained using the property $\chi(-\zeta(\bt))=\chi(\zeta(\bt))^\ast$.
	
	(ii) For the random pulse sequence the initial state is taken to be a coherent state $\vert\varphi\rangle=D(\alpha)\vert 0\rangle$ with $\alpha=3/2$. The corresponding characteristic function has the form
	\begin{equation}
	\chi(\beta)=e^{-\frac{|\beta|^2}{2}}e^{\alpha^\ast\beta-\alpha\beta^\ast}.
	\label{eq:characteristic_coherent}
	\end{equation}
	In the central panel of Fig.~\ref{fig:2}, the green points on the gray surface representing the real part of this characteristic function depict 10\,000 points in reciprocal phase space generated with the random pulse sequence.
	
	(iii) For the linear pulse sequence the initial state is the cat state $\vert\varphi\rangle=[D(\alpha)+D(-\alpha)]\vert 0\rangle/c$ also with $\alpha=3/2$ where the normalization constant reads $c=\sqrt{2[1+\exp(-2|\alpha|^2)]}$. For this state, the characteristic function has the shape
	\begin{align}
	\chi(\beta)=\frac{2 e^{-\frac{|\beta|^2}{2}}}{c^2}\Big[&\cosh\left(\alpha\beta^\ast-\alpha^\ast\beta\right)\nonumber\\
	&+e^{-2|\alpha|^2}\cosh\left(\alpha\beta^\ast+\alpha^\ast\beta\right)\Big].
	\end{align}
	At the bottom of Fig.~\ref{fig:2}, the real part of this characteristic function is again shown in gray, whereas the red and dark gray curves correspond to $\chi(\pm\zeta(\bt))$, respectively.
	
	From the points that are accessible for a direct measurement one can then interpolate the complete characteristic function $\tilde\chi(\beta)$. The quantum state may then be reconstructed by approaching the expansion~\eqref{eq:displacement_expansion} with a sufficiently large sum to approximate the integral according to $\tilde\rho_0=\delta\beta^2\sum_j\tilde\chi(\beta_j) D^\dagger(\beta_j)/\pi$ where the value of $\delta\beta$ depends on the discretization of reciprocal phase space. Alternatively, one can  compute the density matrix elements $\langle n|\tilde\rho_0|m\rangle=\int d^2\beta\,\langle n|D^\dagger(\beta)|m\rangle \tilde{\chi}(\beta)$, as described in the Supplemental Material of Ref.~\cite{Liu2019}. In our examples above, we have considered pure initial states of the harmonic oscillator. In this case, the fidelity of the reconstructed density operator $\tilde{\rho}_0$ with respect to the exact one $\rho_0=\vert\varphi\rangle\langle\varphi\vert$ is simply given by $F=\langle \varphi|\tilde{\rho}_0|\varphi\rangle$, whereas for mixed initial states one has to employ another form~\cite{Jozsa1994}. The characteristic function allows to calculate the fidelity directly without the need to reconstruct the density operator first, according to
	\begin{equation}
	F=\frac{1}{\pi}\int d\beta^2\, \chi(\beta)\tilde{\chi}(\beta)^\ast.
	\end{equation}
	This fidelity is above 99.5\% in all three of the presented cases.

	\subsection{Impact of thermalization on state-reconstruction fidelity}
	We now consider the performance of the three pulse sequences (i)--(iii) in the quantum-state reconstruction for increasing values of the damping rate $\gamma$. For a better comparison, we use the same initial state of the harmonic oscillator for all three pulse sequences, namely a coherent state with amplitude $\alpha=3/2$, whose characteristic function was shown in the central panel of Fig.~\ref{fig:2}. We characterize this performance using the fidelity $F$ introduced above.
	
\begin{figure}[t]
		\includegraphics[width=\linewidth]{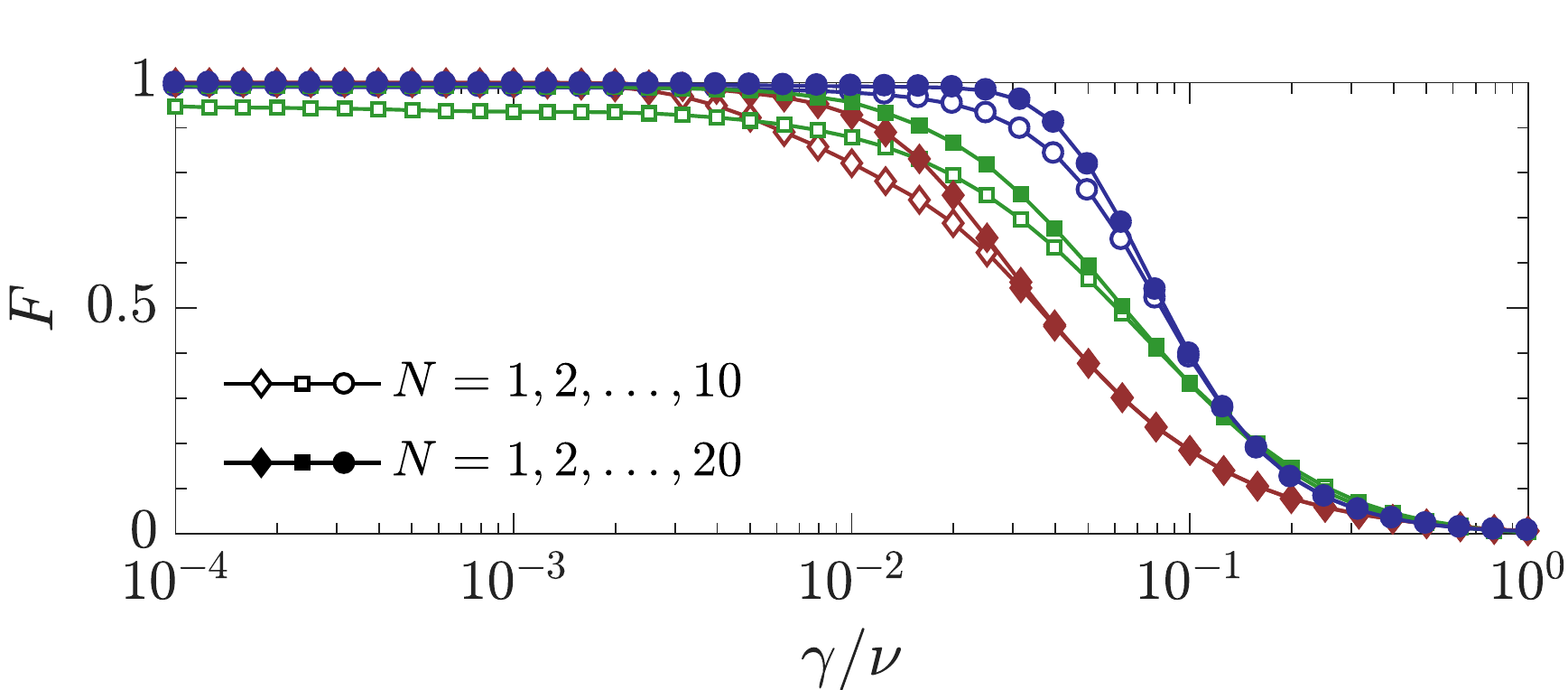}
		\caption{\label{fig:3} Dependence of the reconstruction fidelity $F$ of a coherent state $D(3/2)|0\rangle$ for the three pulses sequences (i)--(iii) on the damping rate $\gamma$: equidistant sequence (i) [blue circles], random sequence (ii) [green squares], and linear sequence (iii) [red diamonds]. Empty markers correspond to a state reconstruction with $N=1,2,\dots,10$, whereas filled markers correspond to $N=1,2,\dots,20$.} 
	\end{figure}

	Figure~\ref{fig:3} shows this fidelity for the interval $\gamma/\nu \in[10^{-4},1]$. For the coupling we have chosen $g/\nu=0.075$ as in the previous figures. Blue, green, and red again correspond to the equidistant, random, and linear pulse sequences, respectively. On the other hand, filled circles show the fidelity when $N=1,2,\dots,20$ is used for an interpolation of the characteristic function, whereas empty circles show the fidelity when only $N=1,2,\dots,10$ is used. In the former case, all three pulse sequences achieve a high-fidelity state reconstruction up to damping rates around $\gamma/\nu=10^{-2}$. For higher damping rates the reconstruction begins to fail and one would have to increase the maximum value of $N$ in order to compensate for the contraction of the accessible curves shown in Fig.~\ref{fig:1}. Since the total pulse-sequence length is the longest for the linear case, the fidelity for this pulse sequence begins to decline first. For the latter case, when only a maximum of ten pulse-sequence blocks are taken into account, the fidelity starts to decline earlier in all three cases. Here, for the random pulse sequence even for low damping rates the accessible points are centered to densely around the origin to allow for a reconstruction with a very high fidelity.
	
	Due to the structure of the pulse sequence with always pairs of equally long free-evolution times, the measurement of the characteristic function inherits a robustness against noise with a long correlation time acting on the probe from pulsed dynamical decoupling~\cite{Uhri2007,Liu2019}. On the other hand, if the probe is subject to uncorrelated noise in the form of pure dephasing with a rate $\Gamma_\text{d}$, the influence on the dynamics can be described by an additional term $\Gamma_\text{d}(\sigma_z\varrho\sigma_z-\varrho)/2$ in the master equation~\eqref{eq:master}. This leads to an exponential decay of the elements of $\varrho$ that are off-diagonal  in the probe degree of freedom. Since it is exactly these off-diagonal elements which allow for the measurement of the characteristic function, their decay can be compensated with a factor $\exp(\Gamma_\text{d}T)$ to nevertheless obtain the correct values of the characteristic function. In an ion-trap setup, noise on the probe could, for example, have its origins in electric or magnetic field fluctuations.

	\section{Conclusion}
	\label{sec:conclusion}
	To summarize, we have presented a method to measure the Wigner characteristic function of an inaccessible harmonic oscillator by coupling it to a controllable probe via a probe-state-dependent force. Outcomes of Pauli measurements of the probe, that was subject to a sequences of pulses that flip its states, are directly related to values of the characteristic function, from which it is possible to reconstruct the initial density operator of the harmonic oscillator.
	
	Our method takes into account thermalization of the harmonic oscillator by a finite-temperature bath. Assuming a Lindblad form of the resulting master equation allows to analytically describe the thermalization during the probe-assisted measurement process, which is conveniently achieved employing superoperators. The key point in the derivation was to transform certain types of superoperators into products of displacements and propagations under the Liouvillian superoperator describing a freely thermalizing harmonic oscillator. This, in turn, made it possible to easily connect probe expectation values with values of the characteristic function.
	
	By varying the shape of the pulse sequence, namely, the number of pulses and the free-evolution times in between them, the value of the characteristic function at different points in reciprocal phase space can be accessed. We have demonstrated this based on three distinct choices for the pulse sequence. The results of the corresponding quantum-state reconstructions show that the pulse-sequence parameters can be adjusted to ensure a high reconstruction fidelity depending on the damping rate.
	
	Furthermore, the pulse sequence is designed in a spin-echo-type fashion, thereby incorporating the features of pulsed dynamical decoupling to compensate noise on the probe with a long correlation time. Also pure dephasing of the probe degree of freedom can be accounted for. The presented method thereby constitutes a robust way to infer the quantum state of a thermalizing harmonic oscillator even in the presence of noise acting on the measurement probe.

	\begin{acknowledgments}
		The authors thank Luigi Giannelli, Simon B. J\"ager, Yiheng Lin, Martin B. Plenio, Kirill Streltsov, Juan Mauricio Torres, and Meichen Yu for helpful discussions and comments. This work was supported by the National Natural Science Foundation of China (Grants No.~11950410494 and No.~11874024).
	\end{acknowledgments}

	\appendix
	\section{Superdisplacement transforms}
	\label{app:displacements}
	In Sec~\ref{sec:propagators}, we relied on the fact that the two superoperators $\cC_{\pm}$ and $\cA_{\pm}$ can be transformed in such a way that their actions are both reduced to that of the Liouvillian superoperator $\cL$ given in Eq.~\eqref{eq:L}. This is conveniently achieved by introducing the superoperator transformations we refer to as superdisplacements. 
	
	\subsection{Symmetric case}
	\label{app:displacements.1}
	In the symmetric case, we want to choose a superdisplacement parameter $\varepsilon$ such that $\cC_\pm =\cD[\mp\varepsilon]\cL\cD[\pm\varepsilon]$ is fulfilled, which can easily be rewritten as $\cD[\pm\varepsilon]\cC_\pm\cD[\mp\varepsilon]=\cL$. Employing $D(\pm\varepsilon)aD^\dagger(\pm\varepsilon)=a\mp\varepsilon$ in every term that occurs when applying $\cC_\pm$ to an arbitrary operator $X$ leads to
	\begin{equation}
	D(\pm\varepsilon)\left[\cC_\pm D^\dagger(\pm\varepsilon)XD(\pm\varepsilon)\right]D^\dagger(\pm\varepsilon)=(\cL\pm\mathcal{Y}) X.
	\end{equation}
	This means we have to choose $\varepsilon$ such that the superoperator $\mathcal{Y}$, with the action
	\begin{equation}
	\mathcal{Y} X=i\left(\varepsilon\tilde{\nu}-\frac{g}{2}\right) [a^\dagger, X]+ i\left(\varepsilon\tilde{\nu}-\frac{g}{2}\right)^\ast [a,X],
	\end{equation}
	vanishes identically. The condition for this simply reads
	\begin{equation}
	\varepsilon=\frac{g}{2\tilde{\nu}},
	\end{equation}
	with the complex frequency $\tilde{\nu}=\nu-i\gamma/2$ we used in the main text. This was the rather simple case and we turn to the more intricate skew case in the following. 
	
	\subsection{Skew case}
	\label{app:displacements.2}
	In the skew case, our aim is to find a supertransformation $\cS[\bx]$ that yields $\cA_\pm =\exp(\xi_1(\xi_3-\xi_2))\cS[\mp\bx]\cL\cS[\pm\bx]-\Gamma\mathcal{I}$, which we used in the main text. In order to achieve this, we start by applying $\cS[\pm\bx]\cA_\pm\cS[\mp\bx]$ to an arbitrary operator $X$. This yields
	\begin{gather}
	\cS[\pm\bx]\cA_\pm\cS[\mp\bx]X=e^{\xi_1(\xi_2-\xi_3)}e^{\pm\xi_1 a}D(\pm\xi_2)\times\nonumber\\
	\left[\cA_\pm D^\dagger(\pm\xi_2)e^{\mp\xi_1 a}Xe^{\pm\xi_1 a}D(\pm\xi_3)\right]D^\dagger(\pm\xi_3)e^{\mp\xi_1 a},
	\label{eq:SAS}
	\end{gather}
	where we have used $\exp(\epsilon a)D(\varsigma)=\exp(\epsilon\varsigma)D(\varsigma)\exp(\epsilon a)$. One can then write out all terms that occur in the above expressions using the two properties
	\begin{gather}
	e^{\epsilon a}D(\varsigma)aD^\dagger(\varsigma) e^{-\epsilon a}=a-\varsigma,\\
	e^{\epsilon a}D(\varsigma)a^\dagger D^\dagger(\varsigma) e^{-\epsilon a}=a^\dagger-\varsigma^\ast+\epsilon,
	\end{gather}
	which shows why we call this superdisplacement skew since $a$ and $a^\dagger$ are displaced differently. We can thereby reshape the superoperator from Eq.~\eqref{eq:SAS} according to
	\begin{equation}
	\cS[\pm\bx]\cA_\pm\cS[\mp\bx]=e^{\xi_1(\xi_2-\xi_3)}(\cL\pm\mathcal{Z}-\Gamma\mathcal{I}).
	\label{eq:SAS2}
	\end{equation}
	The explicit expressions are not very insightful and somewhat lengthy, but we, nevertheless, show them for the sake of completeness. The constant $\Gamma$, which depends on $\bx$, is given by
	\begin{gather}
	\Gamma=\left[i\nu+\frac{(2\nbar+1)\gamma}{2}\right]\xi_2(\xi_2^\ast-\xi_1)-(\nbar+1)\gamma\xi_2(\xi_3^\ast-\xi_1)\nonumber\\
	-\left[i\nu-\frac{(2\nbar+1)\gamma}{2}\right]\xi_3(\xi_3^\ast-\xi_1)-\nbar\gamma\xi_3(\xi_2^\ast-\xi_1)\nonumber\\
	-\frac{ig}{2}(\xi_2+\xi_2^\ast+\xi_3+\xi_3^\ast-2\xi_1)
	\label{eq:GammaApp}
	\end{gather}
	and is identical for both signs in Eq.~\eqref{eq:SAS2}. The superoperator $\mathcal{Z}$, on the other hand, is defined as
	\begin{equation}
	\label{eq:Z}
	\mathcal{Z} X=z_1 a^\dagger X+z_2Xa^\dagger+z_3 aX+z_4 Xa,
	\end{equation}
	with
	\begin{gather}
	z_1=\left[i\nu+\frac{(2\nbar+1)\gamma}{2}\right]\xi_2-\nbar\gamma\xi_3-\frac{ig}{2},\\
	z_2=-(\nbar+1)\gamma\xi_2-\left[i\nu-\frac{(2\nbar+1)\textbf{}}{2}\right]\xi_3-\frac{ig}{2},\\
	z_3=\left[i\nu+\frac{(2\nbar+1)\gamma}{2}\right](\xi_2^\ast-\xi_1)\nonumber\\
	\quad\quad -(\nbar+1)\gamma(\xi_3^\ast-\xi_1)-\frac{ig}{2},\\
	z_4=-\nbar\gamma(\xi_2^\ast-\xi_1)-\left[i\nu-\frac{(2\nbar+1)\gamma}{2}\right](\xi_3^\ast-\xi_1)-\frac{ig}{2}.
	\end{gather}
	We then start by choosing $\xi_2$ and $\xi_3$ such that the first two terms of the superoperator $\mathcal{Z}$ from Eq.~\eqref{eq:Z} vanish, viz., $z_1 a^\dagger X+z_2Xa^\dagger=0$ since they do not include $\xi_1$. This is obviously fulfilled if $z_1=z_2=0$ and the condition for this can be summarized in the form of a linear system of equations that reads
	\begin{equation}
	\begin{pmatrix}
	2\nu-i(2\bar{n}+1)\gamma& 2i\bar{n}\gamma\\
	2i(\bar{n}+1)\gamma & -2\nu-i(2\bar{n}+1)\gamma
	\end{pmatrix}
	\begin{pmatrix}
	\xi_2\\
	\xi_3
	\end{pmatrix}=
	\begin{pmatrix}
	g\\
	g
	\end{pmatrix}.
	\end{equation}
	This system is solved by the two values
	\begin{gather}
	\xi_2=\frac{g}{4|\tilde\nu|^2}\left[2\nu+i(4\bar{n}+1)\gamma\right],\\
	\xi_3=-\frac{g}{4|\tilde\nu|^2}\left[2\nu-i(4\bar{n}+3)\gamma\right],
	\end{gather}
	which are related by $\xi_2-\xi_3=2\varepsilon^\ast$. The condition that the remaining two terms in Eq.~\eqref{eq:Z}, i.e., $z_3aX+z_4Xa$, also vanish reads
	\begin{equation}
	\xi_1= -\frac{ig(2\nbar+1)\gamma}{|\tilde\nu|^2},
	\end{equation}
	since in this case one finds $z_3=z_4=0$. These are the results we summarized in Eq.~\eqref{eq:xi} of the main text. Substituting this choice for $\bx$ into the constant $\Gamma$ given in Eq.~\eqref{eq:GammaApp} yields
	\begin{equation}
	\Gamma=\frac{g^2(2\nbar+1)\gamma}{2|\tilde{\nu}|^2},
	\end{equation}
	which, in fact, is nothing but $ig\xi_1/2$.
	
	Consequently, for this choice of $\bx$ we have guaranteed that the supertransformation~\eqref{eq:SAS2} simply reads $\cS[\pm\bx]\cA_\pm\cS[\mp\bx]=\exp(\xi_1(\xi_2-\xi_3))(\cL-\Gamma\mathcal{I})$ and we can now apply $\cS[\mp\bx]$ from the left and $\cS[\pm\bx]$ from the right to arrive at
	\begin{equation}
	\cA_{\pm}=e^{\xi_1(\xi_3-\xi_2)}\cS[\mp\bx]\cL\cS[\pm\bx]-\Gamma\mathcal{I}.
	\end{equation}
	By employing $\cS[\pm\bx]\cS[\mp\bx]=\exp(\xi_1(\xi_2-\xi_3))\mathcal{I}$ it is straightforward to bring this into the exponential form shown in Eq.~\eqref{eq:asymmetric_displacement}.

	\section{Thermalization of displacement operators}
	\label{app:damped_displacements}
	In several places of the main text, we encountered the action of the free-thermalization propagator $\exp(\cL t)$ on displacement operators. We, therefore, briefly show the derivation of the resulting expression for the sake of completeness. The derivation is conveniently achieved using the time evolution of the Wigner function $W_\varsigma(\alpha,0)$ corresponding to a generic displacement operator $D(\varsigma)$ where the second argument of the Wigner function represents the time. Since displacement operators are intrinsically in symmetric ordering~\cite{Cahill1969a}, one can immediately replace the  annihilation and creation operators by the complex variables $\alpha$ and $\alpha^\ast$, respectively, in order to obtain their corresponding Wigner transform, which thereby simply reads
	\begin{equation}
	\label{eq:Wigner_displacement}
	W_\varsigma(\alpha,0)=\frac{1}{\pi}e^{\varsigma\alpha^\ast-\varsigma^\ast\alpha}.
	\end{equation}
	
	At this point, one can employ the Green's function $G(\alpha,\eta;t)$ of the Fokker-Planck equation for the Wigner function~\cite{Risken1996,Carmichael2002}, i.e., the partial differential equation for the phase-space distribution corresponding to the thermalizing-harmonic-oscillator master equation for the density operator. Explicitly, it has the Gaussian form
	\begin{equation}
	\label{eq:Green}
	G(\alpha,\eta;t)=\frac{1}{\pi\nbar(t)}e^{-\frac{|\alpha-\eta(t)|^2}{\nbar(t)}},
	\end{equation}
	with the two abbreviations $\eta(t)=\eta\exp(-i\nu t-\gamma t/2)$ and $\nbar(t)=(\nbar+1/2)[1-\exp(-\gamma t)]$. The time-evolved Wigner function is then naturally given by the integral of the initial distribution $W_\varsigma(\alpha,0)$ multiplied with the Green's function~\eqref{eq:Green}, namely,
	\begin{align}
	W_\varsigma(\alpha,t)=&\int d^2\eta\,G(\alpha,\eta;t)W_\varsigma(\eta,0)\nonumber\\
	=&e^{\gamma t}e^{-\nbar(t)|\varsigma(t)|^2}W_{\varsigma(t)}(\alpha,0).
	\end{align}
	Here, in order to arrive at the second line, we have simply employed the complex Fourier transform of a Gaussian~\footnote{
		For a constant $\kappa$ with $\Re\{\kappa\}>0$, the complex Fourier transform of a Gaussian reads $\int d^2\varsigma\, \exp(-\kappa\vert\varsigma\vert^2+\alpha\varsigma^\ast-\alpha^\ast\varsigma)=\pi\exp(-|\alpha|^2/\kappa)/\kappa$.
	}
	and defined the shorthand $\varsigma(t)=\varsigma\exp(-i\nu t+\gamma t/2)$. In fact, this is the Wigner function of the displacement operator $D(\varsigma(t))$, apart from the exponential prefactors as can be seen by a comparison to Eq.~\eqref{eq:Wigner_displacement}. 
	
	Again, due to the intrinsic symmetric ordering inherent to displacement operators, one can immediately perform the backtransform from the Wigner function to operators by a simple replacement of phase-space variables with annihilation and creation operators~\cite{Cahill1969a}. We thereby find the expression
	\begin{equation}
	\label{eq:displacement_evolution}
	e^{\cL t}D(\varsigma)=e^{\gamma t}e^{-\nbar(t)|\varsigma(t)|^2}D(\varsigma(t)),
	\end{equation}
	and we see that, contrary to the coherent time-evolution operator, the free-thermalization propagator applied to a displacement operator additionally entails an overall exponential growth as well as a temperature- and time-dependent Gaussian.

	\section{Permuting the free-thermalization propagator and superdisplacements}
	\label{app:cummuting}
	In this appendix, we show how the free-thermalization propagator $\exp(\cL t)$ permutes with the superdisplacement $\cD$ we introduced in Sec.~\ref{sec:propagators}. Some of the identities below rely on results derived in the preceding appendix and have, in fact, already been presented elsewhere in the literature. We nevertheless show the derivations here for self-consistency.
	
	We consider the case of the symmetric superdisplacement $\cD[\epsilon]$. For the sake of brevity, for an arbitrary operator $X$ we define $Y=\exp(\cL t)\cD[\epsilon]X$ in which we can expand $X$ in terms of displacement operators according to
	\begin{equation}
	\label{eq:Y1}
	Y=\frac{1}{\pi}\int d^2\varsigma\,\Tr\{D(\varsigma)X\}e^{\cL t}\cD[\epsilon]D^\dagger(\varsigma),
	\end{equation}
	where we have used the completeness of the displacement operators~\cite{Note1} and the linearity of $\exp(\cL t)\cD[\epsilon]$. 
	
	In the next step, we employ
	\begin{equation}
	\label{eq:DD}
	\cD[\epsilon]D^\dagger(\varsigma)=e^{\epsilon^\ast\varsigma-\epsilon\varsigma^\ast}D^\dagger(\varsigma)
	\end{equation} 
	as well as Eq.~\eqref{eq:displacement_evolution} in order to obtain
	\begin{align}
	Y=\frac{1}{\pi}\int d^2\lambda\,&\Tr\{D(\lambda( t))X\}e^{-\nbar( t)|\lambda|^2}\nonumber\\
	&\times e^{\epsilon( t)^\ast\lambda-\epsilon( t)\lambda^\ast}D^\dagger(\lambda).
	\end{align}
	Here, we have already made the substitution $\lambda=\varsigma(t)$ in the integral and defined $\lambda(t)=\lambda\exp(i\nu t-\gamma t/2)$ as well as $\epsilon( t)=\epsilon\exp(-i\nu t-\gamma t/2)$. Reintroducing the superdisplacement with the displacement parameter $\epsilon(t)$, according to Eq.~\eqref{eq:DD}, allows us to rewrite this expression as
	\begin{equation}
	Y=\cD[\epsilon( t)]\frac{1}{\pi}\int d^2\lambda\left[e^{-\nbar( t)|\lambda|^2}\Tr\{D(\lambda( t))X\}\right]
	D^\dagger(\lambda).
	\end{equation}
	It can be shown that the expression we enclosed in square brackets is, in fact, the time-evolved Wigner characteristic function of the operator $X$, evolved under the action of the differential-operator analogon to $\exp(\cL t)$. We refer to the appendix of Ref.~\cite{Teh2018} for an explicit proof. Therefore we can write
	\begin{equation}
	\label{eq:evolution_characteristic}
	e^{-\nbar( t)|\lambda|^2}\Tr\{D(\lambda( t))X\}=\Tr\left\{D(\lambda)e^{\cL t}X\right\},
	\end{equation}
	which finally yields
	\begin{equation}
	e^{\cL t}\cD[\epsilon]X=\cD[\epsilon( t)]e^{\cL t}X,
	\end{equation}
	after extracting the completeness of the displacement operators~\cite{Note1} again. Since we showed this for an arbitrary operator $X$ the superoperator identity~\eqref{eq:switch_relation_1} used in the main text holds in general. The identity derived here was, e.g., already presented in Ref.~\cite{Deglise2008} without a formal proof.
	
\begin{figure}[b]
		\includegraphics[width=0.75\linewidth]{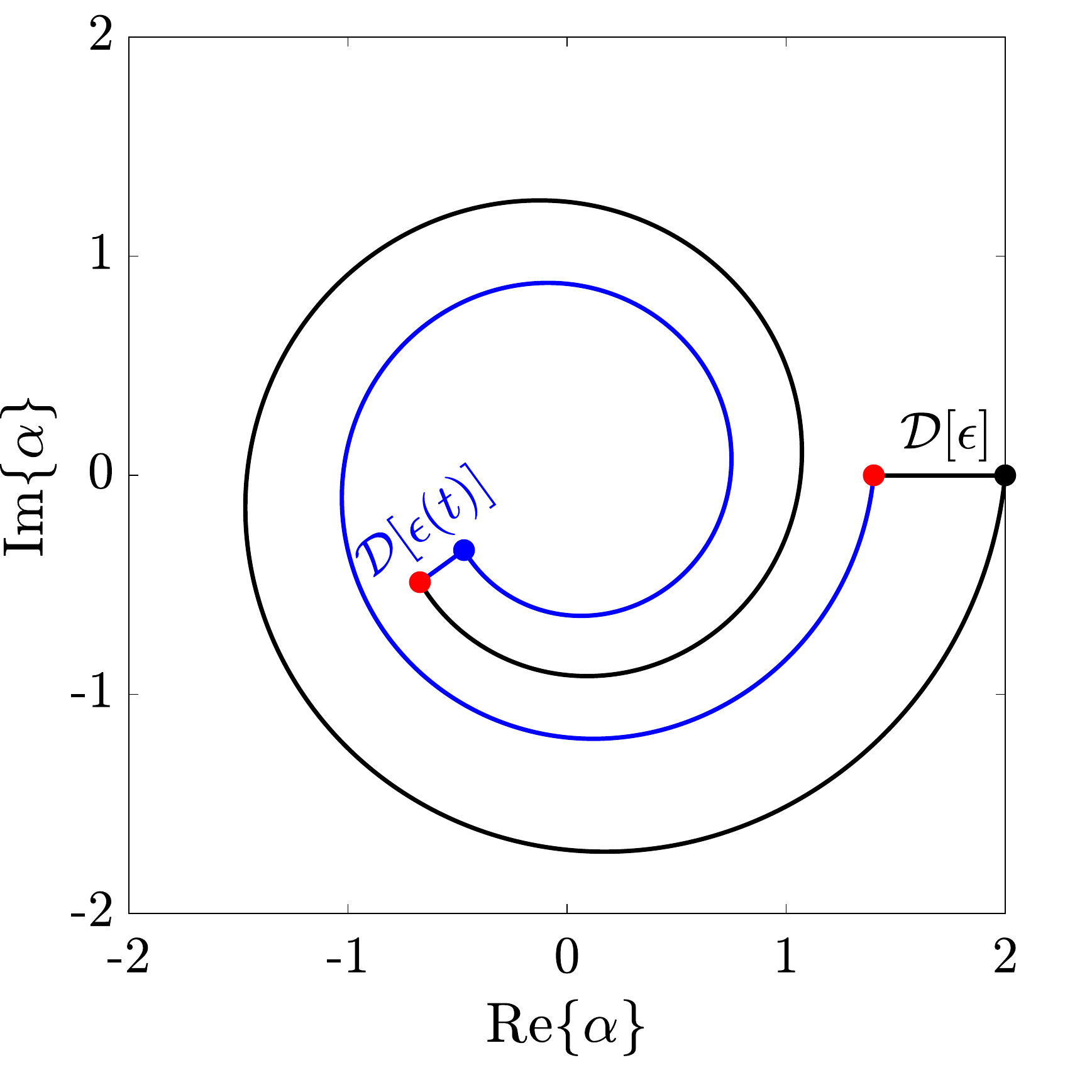}
		\caption{\label{fig:A1}Phase-space trajectories of an initial point $\alpha_0$ (right red dot). Black line: the point is first displaced by $\epsilon$ and then moves along the spiral corresponding to the action of $\exp(\cL t)$. Blue line: the initial point first moves along the spiral (ending in the blue dot) and is then displaced by $\epsilon(t)$. The final points coincide (left red dot).} 
	\end{figure}

	This fact can easily be visualized in phase space, where, apart from thermal diffusion, every point moves along a spiral under the action of the free-thermalization propagator $\exp(\cL t)$. In Fig.~\ref{fig:A1}, we show how an initial point $\alpha_0$ (right red dot) is first displaced by $\cD[\epsilon]$ and then moves along the spiral $(\alpha_0+\epsilon)\exp(-i\nu t-\gamma t/2)$, shown as a black line. However, if the same initial point first moves along the spiral $\alpha_0\exp(-i\nu t-\gamma t/2)$ before it is displaced by $\cD[\epsilon(t)]$, shown as a blue line, the final point (left red dot) is the same.

	\bibliography{article}

\end{document}